\documentclass[10pt]{article}
\usepackage{amsmath}
\usepackage{amsfonts}
\usepackage{amsbsy}
\usepackage{amssymb}
\usepackage{a4wide}
\usepackage{graphicx}

\newcommand{\Nloc}{N_\text{loc}}

\newcommand{\bNloc}{\bar{N}_\text{loc}}

\newcommand{\tNloc}{N^*}

\begin{document}

\title{Response to rebuttal to Comment on Letter ``Excitons in Molecular Aggregates with L\'evy Disorder: Anomalous Localization and Exchange Broadening of Optical Spectra''}
\author{Agnieszka Werpachowska}
\date{April 19, 2012}
\maketitle

\begin{abstract}
The Comment to the Letter ``Excitons in Molecular Aggregates with L\'evy Disorder: Anomalous Localization and Exchange Broadening of Optical Spectra'' appeared in Phys.\,Rev. Lett.~109, 259701 (submitted on November 28, 2011). I prepared the below response to the rebuttal received from the Letter's Authors in the review process. It contains useful comments, which further address the errors in the Letter and other flaws in Authors' understanding of the topic revealed in discussion.
\end{abstract}


\vspace{0.5em}

\noindent In our comment on the Letter ``Excitons in Molecular Aggregates with L\'evy Disorder: Anomalous Localization and Exchange Broadening of Optical Spectra''~\cite{prl10} we have pointed out a couple of errors which led the Authors to incorrect conclusions. The Authors' response to the comment does not invalidate our criticism. On the other hand, it raises some new issues, not mentioned in the Letter. Below, we elaborate some of the points we have raised in the comment and respond to the Authors' new ideas.

We have shown that the title anomalous localization is a consequence of numerical errors. The Authors argued that the universal scaling of the absorption band localisation length with the disorder strength $\sigma$, reported for Gaussian ($\alpha=2$) and Lorentzian disorder ($\alpha=1$) in Ref.~\cite{prb09}, breaks down for $\alpha$-stable disorder distributions with low stability index $\alpha < 1$ (in particular for $\alpha=0.5$ that we will refer to as L\'evy disorder). We have shown that this breakdown results from the incorrect energy range scaling used in the Letter---the constant range $\epsilon \in [-2.1,-1.9]J$ does not adjust to disorder-induced scaling and shifts of the absorption band, like it was assured by using the $\tilde{\epsilon} \in [-0.1,0]$ range (i.e.~$\epsilon$ range scaling with $\sigma$) for the two other disorder types in Ref.~\cite{prb09}. When the energy range is scaled properly, the universality is preserved---see Fig.~1 (inset) in our comment and cf.\ Fig.~3 (inset) in the Letter. The Authors agree that their calculations were incorrect and that the theory works for the L\'evy disorder in the low $\sigma$ regime, where the Letter reports its breakdown. However, they use a new energy range scaling to show that the theory breaks down in the high $\sigma$ regime, where it was supposed to work according to the Letter (see Fig.~3 and related discussion in the Letter). In Sec.~\ref{sec:outliers} of this response we will show that this new breakdown is in fact caused by the high $\sigma$ values and not by the low $\alpha$ index. Hence, it is not specific to systems with disorder described by the distributions with low $\alpha$, but can be observed for any $\alpha$ in high $\sigma$ regime, as we will demonstrate on the example of Gaussian and Lorentzian disorder. Additionally, we note that the new energy scaling used in the Authors' response contains an obvious error.

We have also commented on the breakdown of the half-width at half-maximum (HWHM) scaling with the disorder strength $\sigma$ reported in the Letter. We have shown that if the fitting procedure is performed correctly (the power law should be fitted on a log-log scale or using proper error weights), the HWHM scaling does not break down and follows the relation HWHM $\sim \left(\sigma/|J|\right)^{2\alpha/1+\alpha}$, similarly as in the case of Gaussian and Lorentzian disorder (Fig.~2 in our comment). The Authors do not agree that the HWHM power law should be fitted on a log-log scale, but suggest that it should be done on a linear scale without adjusting error weights. In Sec.~\ref{sec:loglog} we will explain why the log-log scale should be used for this fit. Apart from that, while in the Letter the Authors proposed a new ad-hoc scaling for the HWHM, their response suggests that the scaling simply breaks down for higher $\sigma$ values and no fitting with a continuous function is possible (Fig.~5 of the Authors' response). We comment on it in Sec.~\ref{subsec:Nseg}.

Finally, we have commented that the range of occurrence of the exchange broadening and narrowing in Fig.~2 in the Letter was assessed incorrectly (Fig.~2b in our comment). The Authors agree with this remark.

What we have omitted in our comment, initially considering it an error of minor importance, is that the Authors changed the mathematical definition of heavy-tailedness of $\alpha$-stable distributions to ``decaying slower than Lorentzian'' (i.e.~concerning distributions with $\alpha < 1$). The correct definition says that every $\alpha$-stable distribution with $\alpha < 2$ has heavy tails and generates outliers. Indeed, Ref.~\cite{prb09} deriving the universal localisation length scaling law describes the significant contribution of outliers and the chain segmentation mechanism for Lorentzian disorder (Sec.~III~\cite{prb09}). The Authors argue that we challenge the presence of this mechanism by claiming that the law works also for $\alpha<1$. We will prove in Sec.~\ref{sec:universal} that the universal scaling laws take into account the outliers and the segmentation mechanism. For this reason, they remain universal for all $\alpha$ values and can explain all effects reported in the Letter. Thus, no amendments, like the segmentation law given by~Eq.~(4) in the Letter, are required. Additionally, in Sec.~\ref{subsec:Nseg}, we comment that the segmentation law has problems of its own and does not work in the $\sigma$ regime to which it was applied in the Authors' response.

To summarise, our comment invokes the basic and very important mathematical property of $\alpha$-stable distributions, which will be outlined in Sec.~\ref{sec:universal}---their self-similarity (i.e.~they are stable under convolution). 
It underlies the above scaling laws and their universality observed in molecular aggregates, where the disorder has an $\alpha$-stable distribution and so does---due to the Generalised Central Limit Theorem (GCLT)---the site-averaged disorder experienced by states delocalised by the exchange interaction between molecules. The self-similarity of statistical distributions can have far-reaching physical consequences, leading to a universal description of systems in seemingly completely different environments.
The Letter denies this mathematical property, and that is why we find it necessary to comment on it.

A few corrections have been made to the comment:
\begin{enumerate}
	\item The energy range scaling parameters have been added:
	\begin{quotation}
\noindent ``Using the correct energy range scaling, $\epsilon = \epsilon_b + (5\tilde{\epsilon} +0.23)\sigma^{2/3}$, \dots''
\end{quotation}
	\item The last paragraph has been rephrased:
	\begin{quotation}
\noindent ``The above scaling relations result from the self-similarity of $\alpha$-stable disorder distributions given by the Generalized Central Limit Theorem, $\sigma^\ast=\sigma\Nloc^{1-\alpha/\alpha}$, and thus work for any $\alpha$. The formula for the site-averaged disorder strength carries the information about the type of its distribution (in particular, about the heavy-tailedness for $\alpha<2$). The scaling $\tNloc\sim(J/\sigma)^{\alpha/1+\alpha}$ results from the equilibrium between its typical value experienced by the absorption band states and the typical energetic cost of localizing them. From the universality of $\Nloc$ distribution follows the scaling of $\bNloc$ and $\tNloc$ with equal exponents; together with the above relations, the universality explains also the scaling $\text{HWHM}\sim\sigma(\tNloc)^{1-\alpha/\alpha}$. On the other hand, segmentation (already present in Lorentzian disorder~\cite{prb09}) and localization in potential wells are just microscopic mechanisms realizing this statistical theory.''
\end{quotation}
	\item The lower bound on the localisation length distribution has been additionally mentioned:
		\begin{quotation}
\noindent ``It breaks down only in the limit of small and large $\sigma$ due to the convergence towards $\Nloc=134$ (unperturbed eigenstates) and $\Nloc=1$, respectively.''
\end{quotation}
\end{enumerate}

\section{GCLT as the source of universality of the scaling laws in the molecular aggregates with $\alpha$-stable disorder}
\label{sec:universal}

\begin{figure}[htbp]
\centering
\includegraphics[scale=0.75]{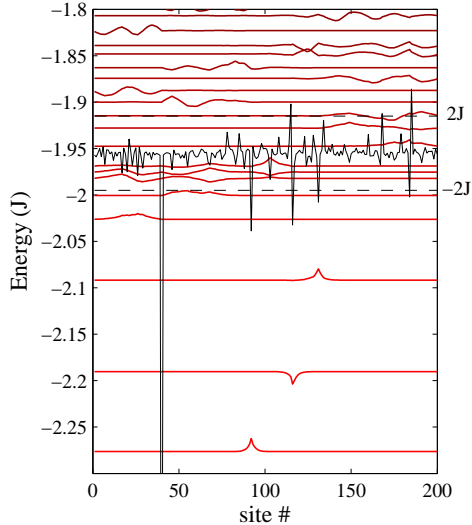}
\caption{Typical set of absorption band states for Lorentzian disorder $\sigma = 0.035J$ with the realisation of disorder energy landscape (rescaled to better display the outliers, $|\epsilon| \ge 2 |J|$).}%
\label{fig:Lorentzscan}%
\end{figure}

In the Letter and in the response to our comment, the Authors state that the scaling laws $N^* \sim \left(\sigma/|J|\right)^{-\alpha/(1+\alpha)}$ and HWHM $\sim \left(\sigma/|J|\right)^{2\alpha/(1+\alpha)}$ do not account for the presence of outliers, and thus cannot work for heavy-tailed $\alpha$-stable disorder distributions. To begin with, we would like to note that the Letter redefines the heavy tails property as ``decaying slower than Lorentzian'' (i.e.~concerning distributions with $\alpha < 1$). In fact, heavy tails (i.e.~tails which are not exponentially bounded and can be asymptotically described by a power law) appear already for distributions with $\alpha < 2$~\cite{nolan2012}. The Authors neglect the fact that Ref.~\cite{prb09} derives the above law for $N^*$ also for Lorentzian disorder, which---according to the correct definition---has heavy tails and generates a significant number of disorder outliers and the chain segmentation (as described in its Sec.~III of Ref.~\cite{prb09} and shown in Fig.~\ref{fig:Lorentzscan}). Apart from this inconsistency, we show below that the main conclusions of the Letter are incorrect---the above laws take into account the outliers and describe not only the ``conventional'' localisation mechanisms, but also the chain segmentation.

\begin{figure}[htbp]
\centering
\includegraphics[scale=0.55]{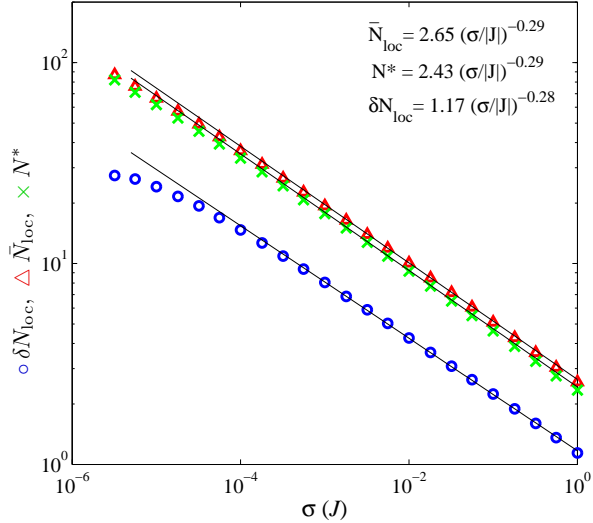}
\caption{Typical $N^*$ and average $\bNloc$ localisation length, as well as its standard deviation $\delta N$ scaling with $\sigma$ for $\alpha=0.4$, using the correct energy range scaling for the absorption band: $\epsilon = \epsilon_b + (5\tilde{\epsilon} + 0.23)\sigma^{4/7}$ (following the approach of Refs.~\cite{prb09,klugkist}). The numerical results (coloured markers) follow closely the scaling law $\left(\sigma/|J|\right)^{-\alpha/(1+\alpha)}$.}%
\label{fig:Nscaling}%
\end{figure}

The relation $\sigma^*(\Nloc) = \sigma \Nloc^{\alpha/(1-\alpha)}$, which is the basis of the scaling law for $N^*$, is a consequence of the self-similarity of $\alpha$-stable distributions (Generalized Central Limit Theorem, GCLT) and is closely related to the existence of their heavy tails. An $\alpha$-stable distribution without heavy tails would have to follow Central Limit Theorem (the finite-variance version of GCLT), leading to the relation $\sigma^*(\Nloc) = \sigma \Nloc^{-1/2}$. Such a distribution is Gaussian with its index of stability $\alpha=2$. For any lower value of $\alpha$, the relation between $\sigma^*$ and $\Nloc$ carries the information about both the existence and asymptotic behaviour of the distribution's heavy tails (disorder outliers) and the behaviour of its core part (moderate disorder, creating the effective potential wells). In other words, $\sigma^*(N)$ is the statistical strength of the average over the sites of all realisations---moderate or extreme---of disorder experienced by a delocalised state. Hence, the $N^*$ in the law derived in Ref.~\cite{prb09} must be the typical localisation length of all states, regardless of the mechanism by which they were localised. It follows that the Letter's statement that the scaling law for $N^*$ does not account for the presence of outliers is incorrect. This is supported by our numerical simulations of $N^*$ for $\alpha < 1$, which show that this law works very well in this regime, even for $\alpha < \frac{1}{2}$ (Fig.~\ref{fig:Nscaling}). The fact that the average localisation length $\bNloc$ and its standard deviation $\delta N$ also scale with the same exponent of $\sigma$ results from the universality of the localisation length distribution, which will be discussed in Sec.~\ref{sec:outliers}.

Consequently, by stating that the above scaling law works for the L\'evy disorder ($\alpha=0.5$) we do not challenge the existence of the chain segmentation by outliers, as implied by the Authors' response. The law results from the above statistical properties of $\alpha$-stable distributions, while the mechanisms of segmentation and localisation on effective potential wells are its microscopic realisations, as described in Ref.~\cite{prb09}.

\begin{figure}[htbp]
\centering
\includegraphics[scale=0.55]{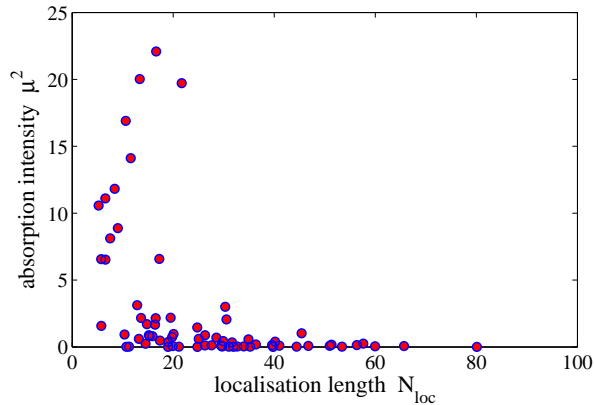}
\caption{An example realisation of absorption intensity of states with different localisation lengths in the Frenkel exciton model with $\alpha$-stable disorder distribution (here for Gaussian disorder with $\sigma=0.35J$).}%
\label{fig:mu2vsNloc}%
\end{figure}

Neither is the GCLT relation for $\sigma^*(N)$ incompatible with the blueshift of the absorption band for $\alpha < 1$. This follows from the properties of states in the considered Frenkel exciton model of a molecular aggregate with disorder in site energies. The absorption band states with shorter localisation lengths are more optically active, while those with longer lengths are mostly optically inactive (Fig.~\ref{fig:mu2vsNloc}). The $\sigma^*(N)$ relation indicates that states with different localisation lengths experience different average disorder strengths. This explains the effective changes of the width and position of the absorption spectrum with $\sigma$. In the case of Gaussian disorder, the optically inactive states experience weaker average disorder $\sigma^* \sim \sigma / \sqrt{\Nloc}$ (exchange narrowing) and thus are spread less away from the band centre $\epsilon = 0$ than the optically active states, pushing the absorption band away from it (red-shift). For L\'evy disorder, they experience larger averaged disorder $\sigma^* \sim \sigma \Nloc$ (exchange broadening) and are spread more away from the band centre than the optically active states, pushing the absorption band towards it (blue-shift). For Lorentzian disorder, all states are spread equally ($\sigma^* = \sigma$) and no shift (or exchange-induced scaling) of the absorption band occurs.

The GCLT relation can also describe the emergence of absorption band peaks made of monomer and dimer states. For disorder distributions with $\alpha < 1$ the shortest states experience smallest disorder and thus are dispersed much less than longer ones, eventually freezing at fixed energy values.

In our comment, and consequently in this response, we agree with the Authors that the chain segmentation is responsible for the blueshift of the absorption band. However, it is not true that this mechanism is present only for $\alpha < 1$. It is present throughout the whole $\alpha$ range of the disorder distributions (outliers can appear sporadically also in Gaussian disorder) and its interplay with the localisation on effective potential wells accounts for the observed absorption band shifts described by the above statistical picture. According to its outcome, we can decipher the microscopic scenarios for different disorder types. For $\alpha > 1$, the optically active states mostly localise in potential wells lowering their potential energy (arising from the disorder) with the growth of $\sigma$, while the effect of the few outliers, squeezing the states on shorter segments and thus pushing up their exchange energy, is weaker. As a result, the red-shift is observed. For $\alpha = 1$ the two effects compensate perfectly. For any $\alpha < 1$, the segmentation effect prevails and the energy of states rises, resulting in the blue-shift of the absorption band, as described in the Letter.

Finally, the HWHM (and FWHM) of the absorption band depends on the typical properties of the absorption band states. Hence, HWHM $\sim \sigma^*(N^*)$, leading to the above HWHM scaling law. This formula also does not neglect the state localisation by outliers, as the scaling law for $N^*$ already takes it into account.

\section{Universality of the localisation length distribution}
\label{sec:outliers}

To demonstrate the breakdown of the universality of the localisation length distribution for high $\sigma$ values, in their response to our comment the Authors propose a new scaling, which includes in the calculations not only the absorption band, but the full absorption spectrum. However, this scaling uses an incorrect exponent $a = 1/3$. We suppose that the proper exponent, describing the absorption band scaling, i.e.~HWHM $\sim \left(\sigma/|J|\right)^{2\alpha/1+\alpha} = \left(\sigma/|J|\right)^{2/3}$, has been confused with that of $N^\star \sim \left(\sigma/|J|\right)^{-\alpha/1+\alpha} = \left(|J|/\sigma\right)^{1/3}$~\cite{prb09,klugkist}. The exponent for the energy range scaling should match the first value, hence $a=2/3$, as used in our comment. We would also like to note that in Fig.~2 in the Authors' response the areas under the probability distributions for different $\sigma$ values should be equal, as should be assured by a proper normalisation.

\begin{figure}[htbp]
\centering
\includegraphics[scale=0.61]{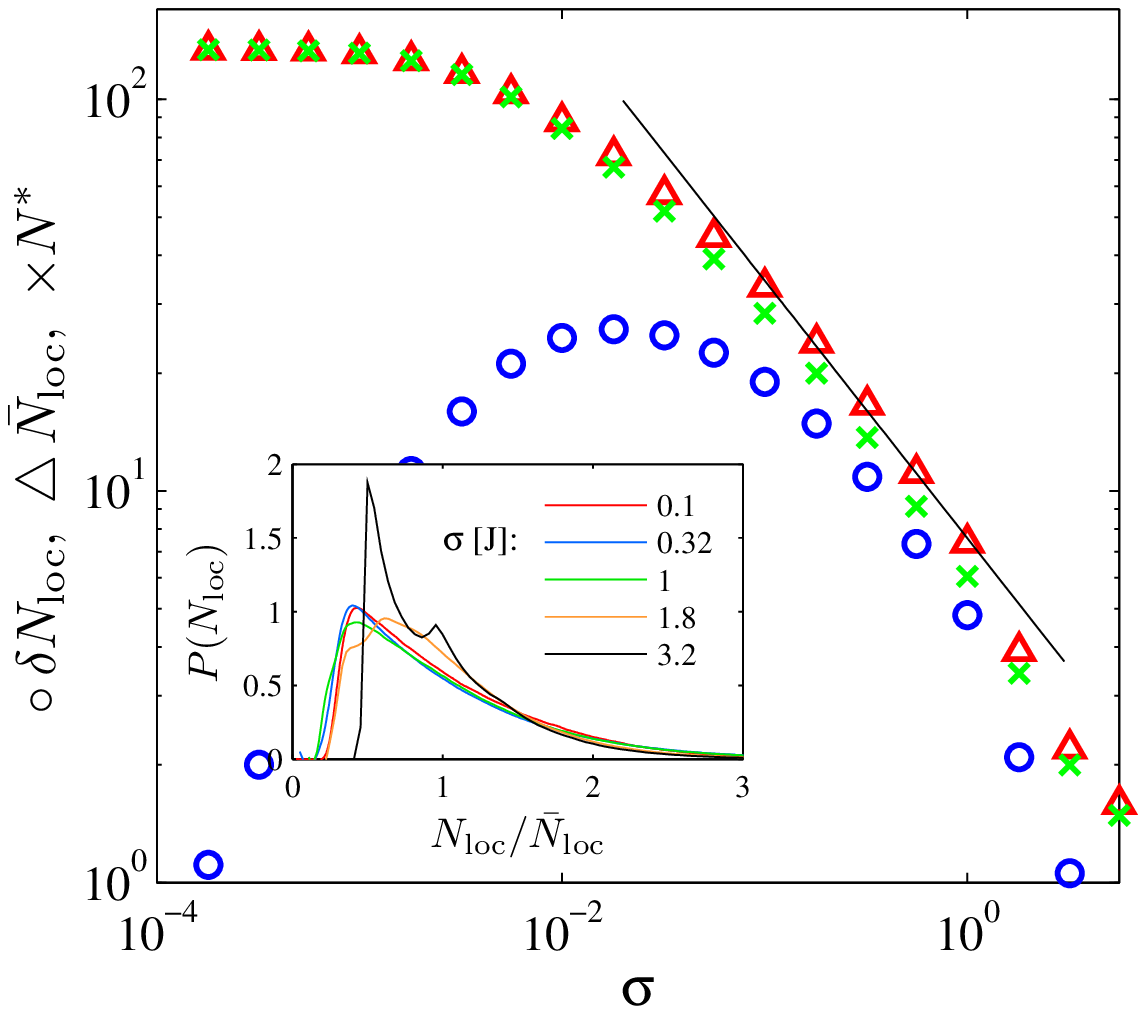}\ \includegraphics[scale=0.61]{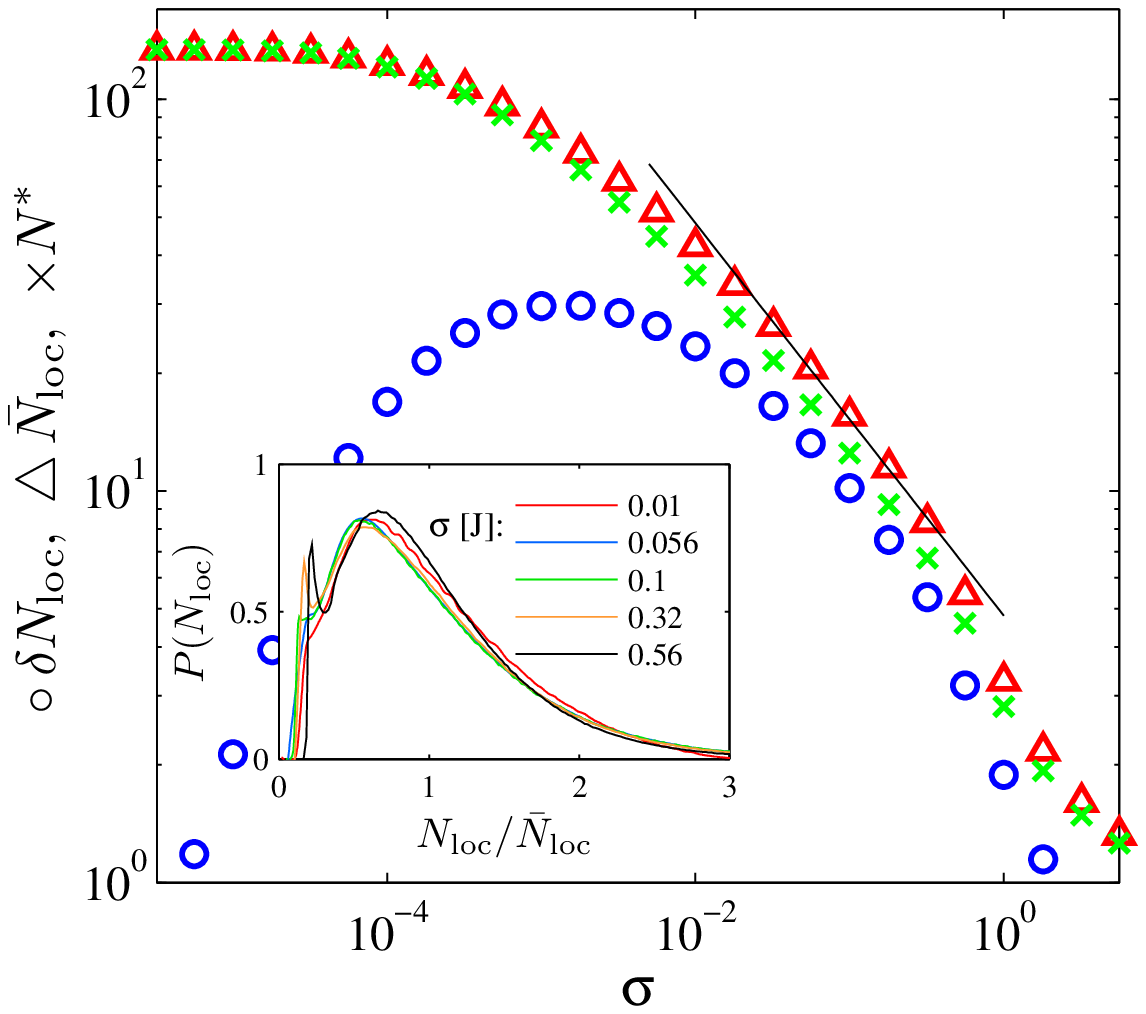}\\
\includegraphics[scale=0.61]{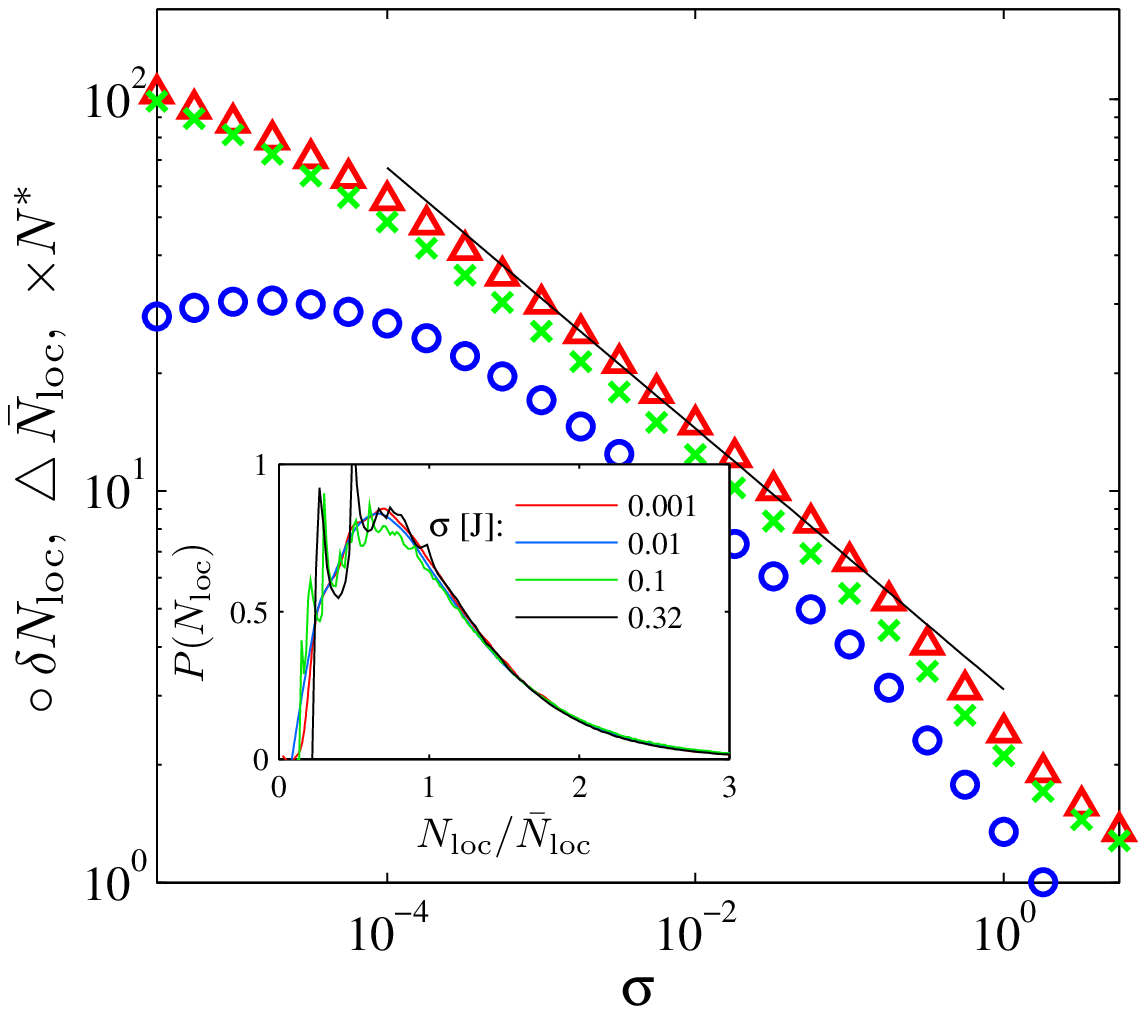}
\caption{Breakdown of the power scaling law and universal distribution of the localisation length (insets) for high $\sigma$ values of Gaussian, Lorentzian and L\'evy disorder (clockwise from upper left) due to the lower bound on localisation length, $\Nloc = 1$. (The scaling law breaks down also for small $\sigma$, where the localisation length is bounded from above by the length of the chain.) Differently than in Fig.~\ref{fig:Nscaling}, we used the new scaling proposed in the Authors' response including full absorption spectrum, but with corrected errors (as described in the text).}%
\label{fig:Nlochighsigma}%
\end{figure}

Supported by the mathematical picture from Sec.~\ref{sec:universal} and correcting the above errors (the exponent and the normalisation), we show that the breakdown of the universality of the localisation length distribution in very strong disorder regime reported in the Authors' response is not the effect of low stability index $\alpha$ of the L\'evy distribution, but of the large $\sigma$ values. In this $\sigma$ regime, the power scaling law breaks down regardless of the disorder type, as demonstrated in Fig.~\ref{fig:Nlochighsigma}. The insets demonstrate that the universality of the localisation length distribution also breaks down (even in Gaussian disorder) if the disorder is sufficiently strong. The simple explanation is that the localisation length is bounded from below by $\Nloc = 1$, and therefore with increasing $\sigma$ and decreasing $\bNloc$, the distribution of $\Nloc / \bNloc$ is ``squeezed'' from below, and fails to be universal for all types of disorder.

\section{Fitting of the HWHM scaling law}
\label{sec:loglog}

Choosing between linear and log-log scale fitting is not a matter of personal preference, but depends on the assumed model (in the sense of constructing statistical estimators of unknown parameters) of random errors causing the deviation of numerical results from theoretical values. In our case, we are estimating the exponent $\alpha$ of a power scaling law $y(x) = a x^\alpha$. If we assume that \emph{absolute} errors are independent from the predictor variables $x_i$,
\[
y_i = a x_i^\alpha + \epsilon_i \ ,
\]
(where $i$ numbers observations of measured variables $y_i$ for different values of $x_i$) one should use a linear scale to obtain an unbiased estimate of exponent $\alpha$~$\cite{linreg}$. On the other hand, if the \emph{relative} errors are independent from $x_i$,
\begin{equation}
y_i = a x_i^\alpha \epsilon_i \ ,
\label{eq:rel-err}
\end{equation}
then by taking the logarithm of both sides we transform this to
\[
\ln y_i = \ln a + \ln \epsilon_i + \alpha \ln x_i \ .
\]
Denoting $\ln a + \ln \epsilon_i$ by $\tilde{\epsilon_i}$, which is a random variable independent of $x_i$, we obtain a linear regression problem. To obtain an unbiased estimate of $\alpha$, we should use a linear scale fit for $(\ln x_i, \ln y_i)$, i.e. a log-log scale fit for $(x_i,y_i)$.

\begin{figure}[htbp]
\centering
\includegraphics[scale=0.55]{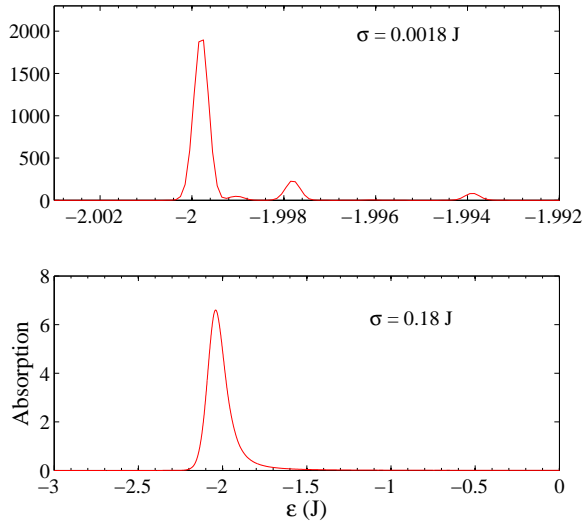}
\caption{Absorption spectrum for the L\'evy disorder of small and large $\sigma$ values.}%
\label{fig:hwhmfit}%
\end{figure}

In our calculations of HWHM, we have used a higher plot resolution to calculate the width of the absorption band for low $\sigma$, where the band is narrower. The relative error arising from this ``measurement error'' was thus independent of $\sigma$, while the absolute error decreased with the decrease of the HWHM. This becomes intuitive when we look at Fig.~\ref{fig:hwhmfit} presenting the absorption spectrum for two $\sigma$ values, 0.0018$J$ and 0.18$J$. The estimated readout error of the width of the second spectrum is of the order of the total width of the first. Additionally, the absolute HWHM error arising from calculating the average absorption band from a sample of a finite (instead of infinite) number of disorder realisations can also be expected to be proportional to the HWHM value. Therefore, the log-log scale fit is the correct one to use when estimating the exponent of the HWHM scaling law.

\section{HWHM scaling}
\label{subsec:Nseg}

In the response to our comment, the Authors use the scaling law for the chain segment length $N_\text{seg}$ distribution proposed in the Letter (Eq.~4) to explain the proposed breakdown of the localisation length scaling for large disorder strengths ($\sigma \geq 0.1J$).

\begin{quotation}
{\small ``It is now important to note that the segmentation mechanism, i.e., the formation of localization segments capped by neighboring outliers in energy, can only be seen if the corresponding segments are typically shorter than the conventional segments. From Eqs. (3) and (4) of the Letter, one can estimate that for $\alpha = 1/2$ the two mechanisms give comparable length scales at $\sigma\approx 0.02$. However, larger values of $\sigma$ are necessary to observe appreciable deviations from conventional localization behavior. At the value $\sigma\approx 0.02$, it is still relatively rare to have segments in the considered energy interval that are sufficiently short to be noticeable (i.e.\ considerably shorter than the conventional localization length $N^*$). As we will show below, only at disorder values of at least $\sigma = 0.1$ or so, will such segments appear frequently within an energy interval near the lower exciton band edge at $E = -2J$. Therefore, we have focused in our Letter on the region with larger values of $\sigma$.''}
\end{quotation}

\begin{figure}[htbp]
\centering
\includegraphics[scale=0.55]{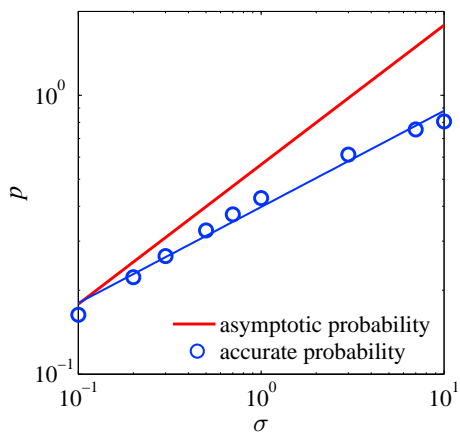}
\caption{Breakdown of the asymptotic scaling relation $p\sim\sigma^{-\alpha}$ for $\alpha=0.5$ in high $\sigma$ regime.}%
\label{fig:4a}%
\end{figure}

\noindent We would like to note that the proposed formula for $N_\text{seg}$ cannot be used all the more so in this regime. This is because it results from the asymptotic scaling law $\sigma^{-\alpha}$ for the outlier occurrence probability in heavy-tailed distributions, which breaks down for $\sigma \geq 0.1J$, as shown in Fig.~\ref{fig:4a}. 

\begin{figure}[htbp]
\centering
\includegraphics[scale=0.75]{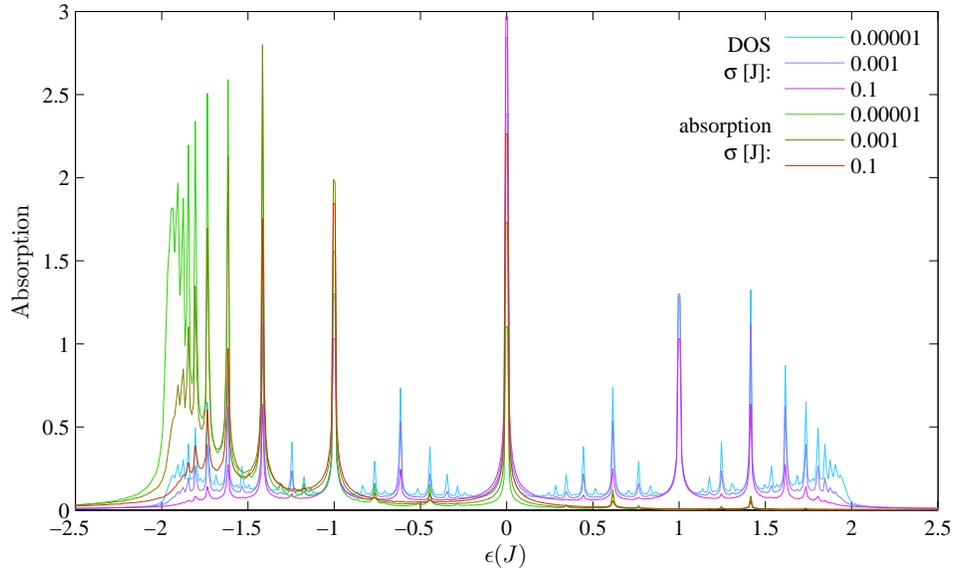}
\caption{Absorption spectrum and the density of states for $\alpha=0.1$.}%
\label{fig:absa0.2}%
\end{figure}

On the other hand, we agree with the Authors in that the perturbative model of HWHM scaling, embodied in the relation HWHM$\sim \sigma^*(N^*)$, cannot be applied for very low $\alpha$ values due to the discretisation of the absorption spectrum by outliers. However, this approach has been adopted by the Authors only in response to our comments, while in the Letter they use the perturbative HWHM model, only with different length scales (based on the incorrect assumption that $N^*$ scaling does not account for the effects of segmentation).  When the discretisation of the density of states starts to split the absorption band into multiple peaks (Fig.~\ref{fig:absa0.2}), this simple perturbative picture does not apply. However, it is not the scaling law which breaks down for the HWHM, but the HWHM itself ceases to be a good characteristic of the absorption band. In summary, we have justified in detail our criticisms and consider that the scientific community will benefit from reading our comment and the Authors' response, including their new ideas.

{\small

}

\end{document}